\documentclass[pra,twocolumn,amsfonts]{revtex4}
\usepackage{amsfonts}
\usepackage{amsmath}
\usepackage{bm}
\usepackage{epsf}

\newcommand{\beq}{\begin{equation}}
\newcommand{\eeq}{\end{equation}}
\newcommand{\beqa}{\begin{eqnarray}}
\newcommand{\eeqa}{\end{eqnarray}}

\newcommand{\ket}[1]{| #1    \rangle }
\newcommand{\bra}[1]{ \langle   #1  | }
\newcommand{\ave}[1]{  \langle #1   \rangle }

\newcommand{\amp }[2]{ \langle #1 |  #2  \rangle }

\newcommand{\rref}[1]{~(\ref{#1})}

\begin{document}

%%%%%%%%%%%%%%%%%%%%%%%%%%%%%%%%%%%%%%%%%%%%%%%%%%%%%%%%%%%%%%%%%%%%%%%%
%%%%%%%%%%%%%%%%%%%% Local Definitions %%%%%%%%%%%%%%%%%%%%%%%%%%%%%%%%%

\newcommand{\suba}[1]{_{_{A_{#1}}}}
\newcommand{\subat}[1]{_{_{\widetilde{A}_{#1}}}}
\newcommand{\subb}[1]{_{_{B_{#1}}}}
\newcommand{\subbt}[1]{_{_{\widetilde{B}_{#1}}}}
\newcommand{\subab}[1]{_{_{A_{#1}B_{#1}}}}
\newcommand{\subabt}[1]{_{_{\widetilde{A}_{#1}\widetilde{B}_{#1}}}}

%%%%%%%%%%%%%%%%%%%%%%%%%%%%%%%%%%%%%%%%%%%%%%%%%%%%%%%%%%%%%%%%%%%%%%%%
%%%%%%%%%%%%%%%%%%%%%%%%%%%%%%%%%%%%%%%%%%%%%%%%%%%%%%%%%%%%%%%%%%%%%%%%

\title[Modewise Entanglement]{Modewise Entanglement of Gaussian States  }

\author{ Alonso Botero }
\email{abotero@uniandes.edu.co} \affiliation{
    Departamento de F\'{\i}sica,
    Universidad de Los Andes,
    Apartado A\'ereo 4976,
    Bogot\'a, Colombia}
\author{ Benni Reznik }
\email{reznik@post.tau.ac.il} \affiliation{ Department of Physics
and Astronomy, Tel-Aviv University, Tel Aviv 69978, Israel.
       }

\date{\today}

\begin{abstract}
\bigskip
We address the decomposition of a multi-mode pure Gaussian state
with respect to a bi-partite division of the modes. For any such
division the state can always be expressed as a product state
involving entangled two-mode squeezed states and single mode local
states at each side. The character of entanglement of the state
can therefore be understood modewise; that is, a given mode on
one side is entangled with only one corresponding mode of the
other, and therefore the total bi-partite entanglement is the sum
of the modewise entanglement. This decomposition is
 generally not applicable to  all mixed Gaussian states. However,
 the result can be
 extended to a special family of ``isotropic" states,
 characterized by a
phase space covariance matrix with a
completely degenerate symplectic spectrum.
\end{abstract}

\maketitle

\section{Introduction}

While the full characterization of bi-partite entanglement for
mixed states is still an open problem, much is known for the case
of pure states. Under a bi-partite division, any pure state may be
written in the Schmidt form
\begin{equation}\label{schmidtform}
    \ket{\psi}_{AB} = \sum_a \sqrt{p_a}
    \ket{\phi_a}_A\ket{\chi_a}_B
\end{equation}
where $\amp{\phi_a}{\phi_b}=\amp{\chi_a}{\chi_b} = \delta_{ab}$,
with unique Schmidt coefficients $\sqrt{p_a}$. The bi-partite
entanglement (entanglement entropy) can then be fixed uniquely by
the asymptotic  yield \cite{distillation} of maximally entangled
states and becomes a function of the Schmidt coefficients only.
Moreover, the Schmidt decomposition  appears to have an
``irreducible" structure: generally speaking, equation
\rref{schmidtform}  cannot be  brought into a simpler form just by
means of local transformations. For instance, a bi-partite system
of $n\times n$ qubits cannot be generally brought to the form of a
product of $n$ entangled pairs under local unitary
transformations.

However, in the context of  Bosonic Channel Capacity,   Holevo and
Werner\cite{holevo01} have shown that a multi-mode\cite{simon94}
Gaussian mixed state can always be purified by enlarging the
system in such way that each normal mode is correlated with a
corresponding single ancillary mode. This procedure achieves a
pure Gaussian state between the system and ancilla in which the
Schmidt decomposition  takes the form of products of bi-partite
two-mode Gaussian states.  Implicit in these results is a general
statement in the converse sense, which  we believe is of
considerable significance for the area of continuous variable
entanglement. The statement is that the  \emph{bi-partite
entanglement of multi-mode Gaussian pure states is in fact
reducible to the product of entangled pairs of single modes}. In
other words, bi-partite entanglement of a Gaussian pure state is
essentially $1 \times 1$ mode Gaussian entanglement.

This result is directly applicable to various problems such as
quantum-optical realizations of quantum information processing
with Gaussian states\cite{wolf}, and the characterization of the
entanglement content of harmonic oscillator chains\cite{plenio}
and bosonic quantum-fields\cite{botero,reznik}. Consider for
instance the vacuum state of a free scalar field, which is
Gaussian.  While one would expect that in this state, the
structure of entanglement  between a given region of space and its
complement would be of a rather complicated nature, such
entanglement  in fact occurs along separate ``channels", with each
member of a set of collective modes in one region correlated with
a corresponding single collective mode of the other.

In this paper we present two different frameworks from which the
modewise decomposition of Gaussian states  can be deduced, and
discuss some of its implications. In the following section, we
present this modewise decomposition in the form of a theorem
applicable to
 arbitrary entangled pure Gaussian states and show how it follows
 from properties of the Schmidt decomposition.
 In section \ref{isotropic} we deal with the case of mixed states.
  Using the correspondence between  correlation matrices and
  Gaussian states, the modewise decomposition implies a
  corresponding decomposition of covariance matrices. We therefore
  show how such a decomposition also holds for a certain class  of
   ``isotropic" Gaussian mixed states, defined from a corresponding
    symmetry of their covariance matrix.

\section{Modewise Decomposition of Pure Gaussian States}

To begin with, suppose a collection of $N$ canonical systems or
``modes"  is partitioned into two sets, i.e., Alice's $A =
\{A_i,\ldots,A_m\}$ and Bob's $B = \{B_1\ldots,B_n\}$, of sizes
$m$ and $n$ respectively.  If the quantum state of the modes is a
pure Gaussian state $\ket{\psi}_{AB}$, the following theorem
characterizes the  entanglement between Alice and Bob:

{\em {Theorem 1:} A Gaussian pure state $\ \ket{\psi}_{AB}$  for
$m + n$ modes $A$ and $B$ may always be written as}
\begin{equation}\label{decomp}
\ket{\psi}_{AB} =
\ket{\widetilde{\psi}_1}\subabt{1}\ket{\widetilde{\psi}_2}\subabt{2}\ldots
\ket{\widetilde{\psi}_s}\subabt{s}\ket{0}\subat{F}\ket{0}\subbt{F}
\end{equation}
{\em for some $s \leq \min(m,n)$, where
$\widetilde{A}=\{\widetilde{A}_1\ldots,\widetilde{A}_m\}$ and
$\widetilde{B}=\{ \widetilde{B}_1,\ldots,\widetilde{B}_n\}$ are
new sets of modes obtained from $A$ and $B$ respectively through
local linear canonical transformations, the states
$\ket{\widetilde{\psi}_k}$ are two-mode squeezed states
\cite{Arvind95} of the form}
\begin{equation}\label{tmss}
\ket{\widetilde{\psi}_i}\subabt{i} = \frac{1}{\sqrt{Z_i}}
\sum_{n}e^{-\frac{1}{2}\beta_i n}\ket{n}\subat{i}\ket{n}\subbt{i}
\, ,
\end{equation}
{\em entangling the modes $\widetilde{A}_k$ and $\widetilde{B}_k$
for $ k\leq s$, and $\ket{0}\subat{F}$ and $\ket{0}\subbt{F} $ are
products of oscillator ground states for the remaining modes in
$\widetilde{A}$ and $\widetilde{B}$ respectively. }

Before proving theorem 1, we first review some facts concerning
the correspondence between Gaussian states and covariance
matrices. Let us represent the canonical variables of a $k$-mode
system by the vector
\begin{equation}
    \eta = \eta_1\oplus\eta_2\oplus\,...\oplus\eta_k \, , \ \end{equation}
where $\eta_i$ is the two component vector $ \eta_i =(q_i,p_i)^T
\, , $ and assume throughout that $\ave{\eta}=0$ for all states
considered. A  generally mixed Gaussian state $\rho$ describing a
system of $k$-modes with $\ave{\eta}=0$ is completely specified by
its covariance matrix (CM), defined as
\begin{equation}\label{covmat}
M = \mathrm{Re}\ave{\eta \, \eta^T} \, .
\end{equation}
A unitary transformation on $\rho$  preserving the Gaussian
character of the state implements a  linear transformation of the
modes $\widetilde{\eta} = S \eta$, known as a symplectic (or
linear canonical) transformation $S \in Sp(2k,\mathbb{R})$. Such a
transformation  preserves the canonical structure of the
commutation relations $[\eta,\eta^T] = i J_{2k}$, where the
$k$-mode {\em symplectic matrix} is given by
$$
J_{2k} =  \bigoplus_{i=1}^{k}J_{2} \, , \ \ \ \ J_2=
\left(%
\begin{array}{cc}
  0 & 1 \\
  -1 & 0 \\
\end{array}%
\right)\, ,
$$
and satisfies $J_{2k}^2  = -\openone_{2k}$.  Hence, a  symplectic
transformation preserves the symplectic matrix under a similarity
transformation, i.e.,
\begin{equation}
\label{simptran} S J_{2k} S^T = J_{2k} \ \Rightarrow\ -
(J_{2k}S)(J_{2k}S^T) = \openone_{2k} \, .
\end{equation}
Under such a transformation, the state $\rho$ is brought to a new
state $\tilde{\rho}$ with CM  $\widetilde{M} = S M S^T$. In
particular, there exist symplectic transformations bringing the CM
to the so-called Williamson normal form  (WNF)
\cite{williamson,will-simon} \label{standia}
\begin{equation}
W =  \lambda_1\openone_2\oplus \lambda_2\openone_2\oplus...\oplus
\lambda_k\openone_2 \, ,
\end{equation}
where $\lambda_i$ are the non-negative eigenvalues of the matrix $
i J_k M$, also known as the {\em symplectic eigenvalues}.
Expressed in  the product Hilbert space corresponding to the new
set of modes $\{\widetilde{\eta}_i\}$ for which $W =
\mathrm{Re}\ave{\widetilde{\eta}\, \widetilde{\eta}^T}$, the
Gaussian state $\rho$ acquires the particularly simple form
\begin{equation}\label{rhogauss}
\rho = \rho_1 \otimes \rho_2 \, \otimes...\otimes \, \rho_k
\end{equation}
where  $\rho_i$ is an oscillator thermal state for the $i$-th mode
\begin{equation}\label{thermstate}
\rho_i = \frac{e^{- \beta_i N_i }}{\textrm{Tr}(e^{- \beta_i N_i})}
= \frac{1}{Z_i}\sum_n e^{-\beta_i n}\ket{n}_i\bra{n}_i \, .
\end{equation}
Here, $\widetilde{N}_i=\widetilde{a}_i^\dag \widetilde{a}_i$ is
the number operator associated with  $\widetilde{a}_i =
(\widetilde{q}_i + i \widetilde{p}_i)/\sqrt{2}$, and $\beta_i$ is
related to the symplectic eigenvalue $\lambda_i$ by $\beta_i =
\ln[(\lambda_i + 1/2)/(\lambda_i - 1/2)]$. Note that as a
consequence of the uncertainty principle, admissible Gaussian
states  satisfy the condition $\forall i\, , \lambda_i \geq
\frac{1}{2}\, $, with  pure Gaussian states when $\forall i\, ,
\lambda_i = \frac{1}{2} $. For $\lambda_i = 1/2$, $\rho_i =
\ket{0}_i\bra{0}_i$ is obtained as the limit of \rref{thermstate}
as $\beta_i \rightarrow \infty$.

We now proceed with the proof of theorem 1. The Schmidt
decomposition \rref{schmidtform} automatically yields the diagonal
form of the partial density matrices for $A$ and $B$:
\begin{equation}
    \rho_A =\sum_a p_a
    \ket{\phi_a} \bra{\phi_a} \, \ \ \ \rho_B =\sum_a p_a
    \ket{\chi_a} \bra{\chi_a}\, ,
\end{equation}
which are seen to be of equal rank and spectrum thus showing that the
$p_a$´s are unique.  The basis states $\ket{\phi_a}_A$ and
$\ket{\chi_a}_B$ are also unique (up to phase factors) for
non-degenerate $p_a$ and otherwise  may be chosen to be elements
of any orthonormal basis spanning the degenerate subspace.  Now,
if $\ket{\psi}_{AB}$ is  Gaussian, then the reduced density
matrices are also  Gaussian. Thus,  $\rho_A$ and $\rho_B$  can be
written in the form \rref{rhogauss}\, in terms of the set of modes
bringing the local covariance matrices into WNF. Suppose that
there are $s$ modes in $A$ and $t$ modes in $B$ with symplectic
eigenvalue $\lambda \neq 1/2$. Since the remaining modes  factor
out from the respective density matrices as projection operators
onto their ground state, we may factor $\ket{\psi}_{AB}$ as
$\ket{\widetilde{\psi}}_{AB}\ket{0}\subat{F}\ket{0}\subbt{F}$
where $\ket{0}\subat{F}$ and $\ket{0}\subbt{F}$ are collective
ground states onto the modes with $\lambda=\frac{1}{2}$ and
$\ket{\widetilde{\psi}}_{AB}$ is the generally entangled state for
the remaining modes,  $\widetilde{A}_1 ...\widetilde{A}_s$ and
$\widetilde{B}_1,..\widetilde{B}_t$.  Concentrate then on
$\ket{\widetilde{\psi}}_{AB}$, the partial density matrices of
which may be written as
$$
\widetilde{\rho}_A = \sum_{\vec{n}_A}\frac{e^{-\vec{\beta}_A \cdot
\vec{n}_A}}{Z^{(A)}} \ket{\vec{n}_A}\bra{\vec{n}_A} \,, \ \ \ \
\widetilde{\rho}_B = \sum_{\vec{n}_B}\frac{e^{-\vec{\beta}_B \cdot
\vec{n}_B}}{Z^{(B)}} \ket{\vec{n}_B}\bra{\vec{n}_B}\, ,
$$
where $\vec{n}_A =\{n\subat{1},...,n\subat{s}\}^T$ and
$\vec{n}_B=\{n\subbt{1},...,n\subbt{t}\}^T$ are $s$ and
$t$-dimensional vectors representing  occupation number
distributions on each side and
$\vec{\beta}_A=\{\beta\subat{1},...,\beta\subat{s}\}^T$ and
$\vec{\beta}_B=\{\beta\subbt{1},...,\beta\subbt{t}\}^T$ represent
the distribution of thermal parameters on each side. Now, by our
previous discussion,  both density matrices have the same rank and
the same eigenvalues. This means that there must exist a
one-to-one pairing between the occupation number distributions
$\vec{n}_A$  and $\vec{n}_B$, and such that
\begin{equation}\label{constraint}
\vec{\beta}_A \cdot \vec{n}_A = \vec{\beta}_B \cdot \vec{n}_B\, .
\end{equation}
We now observe that the pairing $\vec{n}_A \leftrightarrow
\vec{n}_B$ is a {\em homogeneous linear map}, since $\vec{n}_A=0$
and $\vec{n}_B = 0$ are paired (all $\beta's \neq 0$) and
$(\vec{n}_A + \vec{n}_A', \vec{n}_B+\vec{n}_B')$ satisfies
\rref{constraint} if $(\vec{n}_A, \vec{n}_B)$ and
$(\vec{n}_A',\vec{n}_B')$ satisfy \rref{constraint}. However, if a
linear map is one-to-one then the domain and range have the same
dimensions. Thus we see that $s = t$, in other words, the number
of modes in $A$ and $B$ with symplectic eigenvalue different from
$1/2$ are the same. Now, label the modes on each side in ascending
order of $\beta$, so that  $ 0 < \beta\subat{1} \leq
\beta\subat{2}\leq \, ...\leq \beta\subat{s} $, $\beta\subbt{1}
\leq \beta\subbt{2}\leq \, ...\leq \beta\subbt{s}$. Consider first
the case $\vec{n}_A = \{1,0,..0\}^T$, yielding the smallest
non-zero value of $\vec{\beta}\cdot\vec{n}_A$. By construction,
this distribution must be paired with  the smallest non-zero value
of $\vec{\beta}_{B}\cdot\vec{n}_B$, which is (or can be taken to
be in the case of degenerate $\beta\subbt{1}$)  $\vec{n}_B =
\{1,0,..0\}^T$. We thus find that \rref{constraint} has a solution
provided that $ \beta\subat{1} = \beta\subbt{1} \, $ (hence
$\lambda\subat{1} = \lambda\subbt{1}$), and by the linearity
property we find  for any $\vec{n}_A$, the map $n\subat{1}
\rightarrow n\subbt{1}=n\subat{1}$. At this point we can repeat
the procedure but applied to the subspace of the remaining modes,
in other words,  solve for a map between $\vec{n}_A' =
\{0,n\subat{2},..,n\subat{s}\}$ and $\vec{n}_B' =
\{0,n\subbt{2},..,n\subbt{s}\}$ such that $\beta_A\cdot\vec{n}_A'
= \beta_B\cdot\vec{n}_B'$. By a similar argument we find that
$\beta\subat{2} = \beta\subbt{2}$ and $n\subat{2} = n\subbt{2}$.
Iterating the procedure until all the components are exhausted, we
find that the admissible  solutions to \rref{constraint} are
$\vec{n}_A = \vec{n}_B$ (with a freedom of re-ordering the labels
of degenerate modes), provided that  $\vec{\beta}_A =
\vec{\beta}_B$. Reconstructing the Schmidt decomposition of
$\ket{\widetilde{\psi}}_{AB}$ from $\rho_{A}$ and $\rho_{B}$ we
see that
\begin{eqnarray}
\ket{\widetilde{\psi}}_{AB} & = & \frac{1}{Z}\sum_{\vec{n}}
e^{-\frac{1}{2}\vec{\beta}\cdot\vec{n}}\ket{\vec{n}}\ket{\vec{n}} \nonumber \\
& = & \bigotimes_{ i=
1}^{s}\left[\sum_{n}\frac{e^{-\frac{1}{2}\beta_i
n}}{Z_i}\ket{n}\subat{i}\ket{n}\subbt{i}\right]\, .
\end{eqnarray}
Thus, $\ket{\psi}_{AB} =
\ket{\widetilde{\psi}}_{AB}\ket{0}\subat{F}\ket{0}\subbt{F}$ is of
the form (2).

\section{Isotropic Gaussian Mixed States}
\label{isotropic}

Although the previous result may be proved directly from general
features of the Schmidt decomposition, it may also be embedded in
the more general framework of mixed entangled Gaussian states.
Theorem 1 can thus be seen to be a special case of a more general
modewise decomposition theorem for  a certain family of
Gaussian mixed states. To define this family, we shall say that a covariance matrix is
{\em isotropic}  if there
exists a symplectic transformation of the modes $W = S M S^T$ with
$S \in Sp(2k,{\mathbb {R} })$ that brings $M$ to  the form
\begin{equation} \label{isodef}
W = \lambda_0 \openone_{2 k} \, , \ \ \ \   \lambda_0 \geq
\frac{1}{2} \, .
\end{equation}
An {\em isotropic Gaussian state} may thus be defined as a
Gaussian state with an isotropic CM (An example of such a state
would be the thermal state of a set of oscillator modes with
degenerate frequencies). Note that all pure Gaussian states are
isotropic with $\lambda_0 = \frac{1}{2}$ ($\hbar \equiv 1$).

 The
more general theorem is a consequence of the following Lemma
concerning isotropic CMs:

{\em { Lemma 1:} Let $M$ be an isotropic CM for $m+n$ modes $\eta
=\eta_A \oplus \eta_B$ with symplectic eigenvalue $\lambda_0$.
Then there exist local symplectic transformations
$\widetilde{\eta}_A = S_A \eta_A$ and $\widetilde{\eta}_B = S_B
\eta_B$ such that upon appropriate pairing  of the modes, the
covariance matrix  takes the form }
\begin{equation}\label{isoform1}
\widetilde{M} =
\widetilde{M}_{\subabt{1}}\!\oplus\,\widetilde{M}_{\subabt{2}}\oplus\,
...\,\oplus\widetilde{M}_{\subabt{s}}\oplus
\lambda_o\openone_{2(n+m-s)}\, \ \ \  \
\end{equation}
{\em for some $s \leq \min(m,n)$, where
$\widetilde{M}_{\subabt{i}}$ is an isotropic  correlation matrix
for the two-mode sector $\eta\subabt{i} =
\eta\subat{i}\oplus\eta\subbt{i}$ of the form }
\begin{equation}\label{isoform2}
\widetilde{M}_{\subabt{i}} = \left(\begin{array}{cccc}
  \lambda_i & 0 &  \kappa_i & 0 \\
  0 & \lambda_i & 0 & - \kappa_i \\
   \kappa_i & 0 &  \lambda_i  & 0 \\
  0 & - \kappa_i & 0 & \lambda_i \\
\end{array}%
\right) \, , \ \ \ \ \kappa_i^2 =\lambda_i^2-\lambda_0^2\, .
\end{equation}
{\em The diagonal elements $\lambda_i$ in \rref{isoform2} are at
the same time the symplectic eigenvalues of the local CMs $M_A =
\mathrm{Re}\ave{\eta_A\, \eta_A^T}$ and $M_B =
\mathrm{Re}\ave{\eta_B\, \eta_B^T}$ differing from $\lambda_0$,
and the last block in \rref{isoform1} gives the CM for the
remaining modes.}

Given the correspondence between CMs and Gaussian mixed states,
the following extension of Theorem 1 immediately follows:

{\em { Theorem 2:} An isotropic  Gaussian state $\rho^{(0)}_{AB}$
of symplectic eigenvalue $\lambda_0$ for  the $m+n$ modes $A$ and
$B$  may always be written in the form}
\begin{equation}\label{decomp2}
\rho^{(0)} =
\widetilde{\rho}\subabt{1}\otimes\widetilde{\rho}\subabt{2}\otimes...
\otimes\widetilde{\rho}\subabt{s}\otimes\widetilde{\rho}^{(0)}\subat{F}
\otimes\widetilde{\rho}^{(0)}\subbt{F}
\end{equation}
{\em where the new modes $\{\widetilde{A}_i\}$ and
$\{\widetilde{B}_i\}$ are obtained by local symplectic
transformations from $A$ and $B$, $\widetilde{\rho}\subabt{i}$ are
mixed Gaussian two-mode states with CM of the form
\rref{isoform2}, and $\widetilde{\rho}^{(0)}\subat{F}$ and
$\widetilde{\rho}^{(0)}\subbt{F}$ are mixed states for the
remaining modes in $\widetilde{A}$ and $\widetilde{B}$
respectively with diagonal isotropic CM of symplectic eigenvalue
$\lambda_0$. }

To prove Lemma 1,  first perform local symplectic transformations
$\widetilde{\eta}_A\oplus \widetilde{\eta}_B= (S_A\oplus
S_B)\eta_A \oplus\eta_B$  bringing  the local CMs $M_{A}=
\mathrm{Re}\ave{\eta_A\, \eta_A^T}$,
$M_{B}=\mathrm{Re}\ave{\eta_B\, \eta_B^T}$ into  WNF.  The total
CM thus obtained  may be written as
\begin{equation}\label{totcm}
    \widetilde{M}=\mathrm{Re}\ave{\eta\, \eta^T}=\left(%
\begin{array}{cc}
  W_A & \widetilde{K} \\
\widetilde{K}^T & W_B \\
\end{array}%
\right)\, ,
\end{equation}
with $W_A = \bigoplus_{i=1}^{m}\lambda\subat{i}\openone_2 $ and
$W_B = \bigoplus_{i=1}^{n}\lambda\subbt{i}\openone_2$. We next
note a useful fact regarding isotropic CMs. If \rref{isodef} is
satisfied, then
  $M =  \lambda_0 S S^T$ for some symplectic transformation $S$. However, as
is easily verified, $S' = SS^T$ is a symmetric symplectic
transformation. Consequently, \rref{simptran} implies that an
isotropic $k$-mode CM satisfies:
\begin{equation} \label{defdegenm}
-(J M )^2 =  \lambda_0^2  \openone_{2 k}\, .
\end{equation}
Replacing \rref{totcm} into \rref{defdegenm}, and using the fact
that $[W,J] = 0$, the following equations are obtained:
\begin{subequations}\label{rel1}
\begin{eqnarray}
W_A^2  - (J_m \widetilde{K})(J_n \widetilde{K}^T)   & = &
\lambda_0^2
\openone_{2m} \label{rel1a}\\
W_B^2  - (J_n \widetilde{K}^T)(J_m \widetilde{K} )  & = &
\lambda_0^2 \openone_{2n}
\label{rel1b}\, \\
-W_A \widetilde{K}  + J_m \widetilde{K} J_n W_B   & = & 0
\label{rel1c}\, .
\end{eqnarray}
\end{subequations}
Consider then a $2 \times 2$ sub-block $\widetilde{K}_{ij} \equiv
\ave{\eta\subat{i} \eta\subbt{j}^T}$ of $\widetilde{K}$. Since
$W_A$ and $W_B$ are diagonal, from equation \rref{rel1c} we find
that
\begin{equation}
\lambda\subat{i}\widetilde{K}_{ij} =\lambda\subbt{j} J_2
\widetilde{K}_{ij} J_2 \, .
\end{equation}
It is not hard to verify that unless $\lambda\subat{i} =
\lambda\subbt{j}$, this equation has no solution for
$\widetilde{K}_{ij}$ other than $\widetilde{K}_{ij}=0$. Thus we
find that modes in $A$ and modes $B$ with different symplectic
eigenvalues are uncorrelated

Next, let $\widetilde{\eta}\subat{\lambda}$ and
$\widetilde{\eta}\subbt{\lambda}$ stand for the  modes in $A$ and
$B$ with the same local symplectic eigenvalue $\lambda$, and group
the modes according to their eigenvalues so that $\widetilde{M}$
takes the Jordan form $\widetilde{M} = \bigoplus_\lambda
\widetilde{M}_\lambda$ where each block $\widetilde{M}_\lambda$ is
the CM for the degenerate eigenmodes
$\widetilde{\eta}\subat{\lambda}\oplus\widetilde{\eta}\subbt{\lambda}$.
Concentrating on a given $\lambda$, assume that $g_A$ and $g_B$
are the degeneracies of $\lambda$ in the symplectic spectra of
$W_A$ and $W_B$ respectively, so that $\widetilde{M}_\lambda$ may
be written as
\begin{equation}\label{degcm}
    \widetilde{M}_\lambda=\left(%
\begin{array}{cc}
  \lambda \openone_{2 g_A} & \widetilde{K}_\lambda \\
\widetilde{K}_\lambda^T & \lambda \openone_{2 g_B}\\
\end{array}%
\right)\, .
\end{equation}
Now note that $\widetilde{M}_\lambda$ is also an isotropic CM with
symplectic eigenvalue $\lambda_o$. Substituting \rref{degcm} into
\rref{defdegenm} therefore yields
\begin{subequations}\label{rel2}
\begin{eqnarray}
J_{2g_A} \widetilde{K}_\lambda J_{2g_B} \widetilde{K}_\lambda^T
& = & (\lambda^2 -
\lambda_o^2)\openone_{2 g_A}\label{rel2a} \\
J_{2g_B} \widetilde{K}_\lambda^T J_{2g_A} \widetilde{K}_\lambda
& = & (\lambda^2 - \lambda_o^2)
\openone_{2 g_B} \label{rel2b} \\
J_{2g_A} \widetilde{K}_\lambda J_{2g_B} &=& \widetilde{K}_\lambda
\label{rel2c} \, .
\end{eqnarray}
\end{subequations}
Taking the trace of equations \rref{rel2a} and \rref{rel2b} and
using the cyclic property of the trace $\textrm{Tr}[J_{2g_A}
K_\lambda J_{2g_B} K_\lambda^T  ] = \textrm{Tr}[ J_{2g_B}
K_\lambda^T J_{2g_A} K_\lambda  ]$, we obtain that $ (\lambda^2 -
\lambda_o^2)(g_A - g_B) = 0 \, . $ Thus we see that the respective
degeneracies of the symplectic eigenvalue $\lambda$ in the local
covariance matrices $M_A$ and $M_B$ must be the same for $\lambda
\neq \lambda_0$.  In such a case, let $g_A = g_B = g$ and from
equations \rref{rel2} deduce that
\begin{subequations}\label{rel3}
\begin{eqnarray}
\widetilde{K}_\lambda^T J_{2g} \widetilde{K}_\lambda   =
\widetilde{K}_\lambda J_{2g} \widetilde{K}_\lambda^T      & = &
-(\lambda^2 - \lambda_o^2)J_{2g}\label{rel3a} \\
J_{2g} \widetilde{K}_\lambda J_{2g} \widetilde{K}_\lambda^T &=&
\widetilde{K}_\lambda \widetilde{K}_\lambda^T \label{rel3c} \, .
\end{eqnarray}
\end{subequations}
Next, define a matrix $\beta = \oplus_{i}^{g} \sigma_3$, where
$\sigma_3$ is the standard Pauli matrix, and hence satisfying
$\beta^2 = \openone_{2g}$, $\beta^T = \beta$, $\{\beta,J_{2g}\} =
0$,  and $\beta J_{2g} \beta = - J_{2g}$. Re-expressing $K$ in
terms of some other $2 g \times 2 g$ matrix $O_{\lambda}$ as
\begin{equation}
\widetilde{K}_{\lambda}= \sqrt{\lambda^2 - \lambda_0^2}\,
O_{\lambda}\, \beta \,
\end{equation}
and substituting into   \rref{rel3} we find that  $O_{\lambda}$
must satisfy
\begin{equation}
O_{\lambda} J_{2g} O_{\lambda}^T = J_{2g}\, ,\ \ \ \
O_{\lambda}^T O_{\lambda} = \openone_{2 g}\, ;
\end{equation}
in other words, $O_{\lambda}$ must be an orthogonal symplectic
transformation. One can then perform a one-sided symplectic
transformation, say $O_{\lambda}^T$ on the $A$-modes, leaving the
local CMs invariant in $\widetilde{M}_\lambda$ and  bringing
$\widetilde{K}_\lambda$ into the block diagonal form, i.e.,
\begin{equation}
\widetilde{K}'_\lambda = O_\lambda^T \widetilde{K}_\lambda =
\sqrt{\lambda^2 - \lambda_0^2}\bigoplus_{i}^{g} \sigma_3 \, .
\end{equation}
In this way, we achieve a pair-wise decomposition of the
degenerate subspace itself, where each pair has a covariance
matrix of the form (8). Finally, we note that for the degenerate
subspace associated with $\lambda =\lambda_o$, in which the
degeneracies on each side are not restricted to be the same,
equations \rref{rel3} imply that $\widetilde{K}_{\lambda_o}
\widetilde{K}_{\lambda_o}^T =0 \Rightarrow
\widetilde{K}_{\lambda_o}=0$. Therefore,  local modes with
symplectic eigenvalue $\lambda_o$ decouple, as expected from the
pure case $\lambda_o = \frac{1}{2}$.

To conclude with, we note that it is known that  for a  $1 \times
1$-mode  Gaussian mixed state, the
   Peres-Horodecki partial transpose
criterion\cite{peres-horodecki} is both necessary and sufficient
\cite{simon} for entanglement and hence for
distillability\cite{gaussian-distillable}. In the present case of
a two mode CM of the form \rref{isoform2}, the partial transpose
criterion implies that the state is entangled iff $\lambda_i >
\lambda_0^2 + \frac{1}{4}$. Consequently, an isotropic Gaussian
state is entangled and distillable iff at least one of the
symplectic eigenvalues of its local CMs satisfies this condition.

\section{Acknowledgments}

We give special thanks to Michael Wolf for bringing to our
 attention reference \cite{holevo01}. We would also like to
 thank Sandu Popescu and Noah Linden for very helpful discussions,
 and  the European Science Foundation for support at the 2002
 ESF Meeting on Foundations of Quantum Mechanics and Quantum
 Information, University of Bristol. A.B. acknowledges support
  from Colciencias (contract RC-425-00) and the Summer Institute
   on Foundations of Quantum Mechanics at
   The University of South Carolina. B.R. acknowledges grant
   62/01-1 of the Israel Science Foundation, established by the
    Israel Academy of Sciences and Humanities and the Israel MOD
Research and Technology Unit.


\begin{thebibliography}{99}



\bibitem{distillation} C. H. Bennett, H. Bernstein, S. Popescu, and B.
Schumacher, Phys. Rev. A {\bf 53}, 2046 (1996). S. Popescu and D.
Rohrlich, Phys. Rev. {A} {\bf 56}, 3319 (1997).

\bibitem{holevo01} A. S. Holevo and R. F. Werner, Phys. Rev. {A} {\bf 63}, 032312 (2001).

\bibitem{simon94}
    R. Simon, N. Mukunda, B. Dutta,
%  QUANTUM-NOISE MATRIX FOR MULTIMODE SYSTEMS - U(N) INVARIANCE,
%     SQUEEZING, AND NORMAL FORMS
 Phys. Rev. A {\bf 49} 1567 (1994).




\bibitem{wolf} M.M. Wolf, J. Eisert, M.B. Plenio, {\em
The entangling power of passive optical elements} quant-ph/020617.


\bibitem{plenio} K Audenaert, J. Eisert M. B. Plenio and R. F.
Werner, Phys. Rev. A {\bf 66}, 042327  (2002).

\bibitem{botero} A. Botero and B. Reznik, In preparation.

\bibitem{reznik} B. Renzik {\em Distillation of vacuum entanglement},
quant-ph/0008006, quant-ph/0212044.

\bibitem{Arvind95}
Arvind, B. Dutta, N. Mukunda and R. Simon,
%     2-MODE QUANTUM-SYSTEMS - INVARIANT CLASSIFICATION OF SQUEEZING
%     TRANSFORMATIONS AND SQUEEZED STATES
 Phys. Rev. A {\bf52}, 1609 (1995)

\bibitem{williamson} J. Williamson, Amer. J. Math. {\bf58}, 141 (1936).

\bibitem{will-simon} R. Simon, E.C.G. Sudarshan and N. Mukunda,
Phys. Rev. A {\bf 36} 3868 (1987).

\bibitem{peres-horodecki}
A. Peres, Phys. Rev. Lett. {\bf 77}, 1413 (1996). M. Horodecki, P.
Horodecki and R. Horodecki, Phys. Lett. A {\bf 223}, 1 (1996).



\bibitem{simon} R. Simon, Phys. Rev. Lett. {\bf84}, 2726 (2000)

\bibitem{gaussian-distillable}
% Entanglement purification of Gaussian continuous variable quantum states
L. M. Duan, G. Giedke G, J. I. Cirac and P. Zoller P, Phys. Rev.
Lett. {\bf 84} 4002 (2000).





-----------------------------------------------

%\bibitem{Huang94}
%HUANG H, AGARWAL GS
%     GENERAL LINEAR TRANSFORMATIONS AND ENTANGLED STATES
%     PHYS REV A 49 (1): 52-60 JAN 1994

%\bibitem{Simon88}
%SIMON R, SUDARSHAN ECG, MUKUNDA N
%     GAUSSIAN PURE STATES IN QUANTUM-MECHANICS AND THE SYMPLECTIC
%     GROUP
%     PHYS REV A 37 (8): 3028-3038 APR 15 1988




\end{thebibliography}
\end{document}